\documentclass[twocolumn]{aastex6}
\usepackage{natbib}
\usepackage{color}

\usepackage{color}
\usepackage{hyperref} 

\def \bea {\begin{eqnarray}}
\def \ena {\end{eqnarray}}                  
\def \bee {\begin{equation}}
\def \ene {\end{equation}}
\def    \simlt  {\lower.5ex\hbox{$\; \buildrel < \over \sim \;$}}
\def    \simgt  {\lower.5ex\hbox{$\; \buildrel > \over \sim \;$}}

\newcommand     \mum    {\,\mu{\rm m}}  

\def	\cm		{\,{\rm {cm}}}

\def	\km		{\,{\rm {km}}}

\def	\erg		{\,{\rm {erg}}}

\def	\g		{\,{\rm g}}

\def	\K		{\,{\rm K}}

\def	\s		{\,{\rm s}}

\def	\H		{\rm H}

\def    \apjl  		{\rm {ApJL}}

\def	\gas		{\rm {gas}}

\shorttitle{Spinup and Disruption of interstellar asteroids}
\shortauthors{Hoang et al.}

\begin{document}
\title{Spinup and Disruption of Interstellar Asteroids by Mechanical Torques, and Implications for 1I/2017 U1 (`Oumuamua)}

\author{Thiem Hoang}
\affil{Korea Astronomy and Space Science Institute, Daejeon 34055, Korea, email: thiemhoang@kasi.re.kr}
\affil{Korea University of Science and Technology, Daejeon, 34113, Korea}

\author{Abraham Loeb}
\affil{Harvard-Smithsonian Center for Astrophysics, 60 Garden Street, Cambridge, MA, USA}

\author{A. Lazarian}
\affil{Astronomy Department, University of Wisconsin, Madison, WI 53706, USA}

\author{Jungyeon Cho}
\affil{Department of Astronomy and Space Science, Chungnam National University, Daejeon, Korea}

\begin{abstract}
The discovery of the first interstellar asteroid, 1I/2017 U1 (`Oumuamua), has opened a new era for research on interstellar objects. In this paper, we study the rotational dynamics of interstellar asteroids (ISAs) of irregular shapes moving through the interstellar gas. We find that regular mechanical torques resulting from the bombardment of gas flow on the irregular body could be important for the dynamics and destruction of ISAs. Mechanical torques can spin up the ISA, resulting in the breakup of the original ISA into small binary asteroids when the rotation rate exceeds the critical frequency. We find that the breakup timescale is short for ISAs of highly irregular shapes and low tensile strength. We apply our results to the first observed ISA, `Oumuamua, and suggest that its extreme elongated shape may originate from a reassembly of the binary fragments due to gravity along its journey in the interstellar medium. The tumbling of `Oumuamua could have been induced by rotational disruption due to mechanical torques. Finally, we discuss the survival possibility of high-velocity asteroids presumably formed by tidal disruption of planetary systems by the black hole at the Galactic center.
\end{abstract}

\keywords{asteroids: individual (1I/2017 U1 (`Oumuamua)) — meteorites, meteors, meteoroids}

\section{Introduction}\label{sec:intro}
The detection of the first interstellar object, 1I/2017 U1 (`Oumuamua) by the Pan-STARRS survey \citep{2017MPEC....U..181B} implies an abundance population of similarly interstellar objects (\citealt{Meech:2017hu}; \citealt{2018ApJ...855L..10D}). The elongated shape with an extreme axial ratio of $\gtrsim 5:1$ of `Oumuamua is mysterious (\citealt{Fraser:2018dg}; see also \citealt{2017ApJ...850L..36J} and \citealt{2015MNRAS.454..593B}). \cite{2017ApJ...851L..38B} and \cite{Gaidos:2017wj} suggested that it may be an elongated object or a contact binary, while others speculate that the bizarre shape might be formed by violent events, such as collisions during the planet formation stage. The discovery of the first ISA opens a new era of research on interstellar objects.

Rotational dynamics is critically important for understanding the formation and evolution of asteroids. For solar system asteroids (SSAs), it is widely believed that the rotation of small asteroids is excited by Yarkovsky-O’Keefe-Radzievskii-Paddack (YORP; \citealt{Rubincam:2000fg}; see also \citealt{2003Natur.425..131B}), while larger ones are driven by random collisions with asteroids in the asteroid belt (see \citealt{2006AREPS..34..157B} for a review). Tidal encounters between the asteroid and a planet system can disrupt large SSAs. 

For ISAs such as `Oumuamua, the YORP effect is unlikely important because this ISA does not revolve around the Sun as SSAs. Moreover, collisions of ISAs are expected to be much less frequent due to the low density of the ISAs (see e.g., \citealt{2018ApJ...852L..27F}). Therefore, to understand the origin and evolution (in shape and size) of ISAs, it is necessary to study physical processes that can affect the dynamics of ISAs in the interplanetary and interstellar medium (ISM).

Photometric observations reveal that SSAs have different shapes, from highly irregular shapes to spheroidal shapes. Thus, we expect ISAs to also have a variety of irregular shapes. Because of their motion through the gas, ISAs should experience regular mechanical torques in the same way as an helical dust grain or a windmill gets spun-up by the gas flow. The original idea of mechanical torques for irregular grain shapes was introduced by \cite{2007ApJ...669L..77L} where the authors found that the gas flow can produce strong regular mechanical torques when interacting with the helical grain. Numerical calculations in \cite{2016MNRAS.457.1958D} demonstrated the spinup effect by mechanical torques for dust grains of Gaussian random shapes. Recently, \cite{2018ApJ...852..129H} quantified the mechanical torques for realistic grains and found that for highly irregular shapes, mechanical torques are efficient in spinning-up grains to suprathermal rotation as well as aligning grains with magnetic fields.

Due to the increase in mechanical torques with the object's surface area, an ISA is expected to experience large mechanical torques. As a result, ISAs would be spun-up and disrupted when the rotation speed exceeds the critical threshold determined by the maximum tensile strength of the material. Such a rotational disruption would be an important process that breaks an ISA into small fragments in the ISM.

The structure of the paper is as follows. In Section \ref{sec:spinup} we calculate the rotation velocity achieved by mechanical torques for ISAs. In Section \ref{sec:breakup} we discuss the critical speed and breakup timescale of ISAs by mechanical torques. These results are applied to ISAs in Section \ref{sec:discuss}. Finally, \ref{sec:sum} summarizes our main results.

\section{ISA Spinup by Mechanical torques and Impulsive Torques}\label{sec:spinup}
Rotational dynamics is important for understanding the wide range of sizes and rotation periods of small SSAs, as well as the maximum observed rotation rate of large SSAs (\citealt{2000Icar..148...12P}, see also \citealt{2012AJ....143...66J}). We will first study the spinup of IASs by mechanical torques.

\subsection{Spinup by mechanical torques}
\cite{2007ApJ...669L..77L} showed that mechanical torques due to the reflection a gas flow onto a helical grain can spin-up dust grains to suprathermal rotation, i.e., rotating with velocity larger than the thermal velocity. \cite{2018ApJ...852..129H} found that the efficiency of spinup by mechanical torques depends on the irregularity of grain shapes, such that highly irregular shapes can be spun-up to suprathermal rotation, while axisymmetric grains experience negligible spinup. 

SSAs have different shapes, from highly irregular shapes to spheroidal shapes, and a number of SSAs exhibit simple shapes with large planar facets and sharp edges (\citealt{Torppa:2003ii}), such as 2100 Ra-Shalom, 43 Ariadne, and 694 Ekard (see \citealt{2009ApJ...699L..13D}).

Therefore, to account for a wide range of shapes of ISAs, we assume that the asteroid surface can be represented by $N_{\rm fc}$ facets having random orientations. Thus, an axisymmetric spheroid has $N_{\rm fc}\rightarrow \infty$, and the helical grain shape in \cite{2007ApJ...669L..77L} has $N_{\rm fc}=1$. Let $R$ be the effective size of the irregular asteroid with mass density $\rho$. The inertia moment is $I_{i}=8\pi \rho R^{5}\alpha_{i}/15$ where $\alpha_{i}$ are the dimensionless parameters of unity with $\alpha_{i}=1$ for spherical asteroids.

We consider an ISA drifting at speed $v_{d}$ through a gas of hydrogen density $n_{\H}$ and temperature $T_{\gas}$. When a stream of gas particles hit the asteroid surface, each facet can act as a mirror, acquires an amount of the momentum due to reflection of the gas flow that provides a random contribution to the total torque. Following \cite{2018ApJ...852..129H}, the mechanical torque (MAT) due to specular reflection by a facet is approximately given by
\bea
\delta \Gamma_{\rm MAT}\sim \left(n_{\H} \frac{4\pi R^2}{N_{\rm fc}} v_{d}\right) \gamma_r \left( m_{\H} v_{d} R\right),\label{eq:dGamma}
\ena
where $\gamma_r$ is the reflection coefficient, and the grain surface area is $4\pi R$, $n_{\rm H}$ is the gas number density. 

The net torque from $N_{\rm fc}$ facets can be calculated using the random walk formula:
\bea
\Gamma_{\rm MAT} &\sim & \delta \Gamma_{\rm MAT} \sqrt{\gamma_e N_{\rm fc}} \nonumber \\
&\sim& 4 \pi \gamma_r \sqrt{\gamma_e} n_{\H} m_{\H}  v_{d}^2  \frac{R^3}{\sqrt{N_{\rm fc}}}, \label{eq:GamMAT}
\ena
where $\gamma_e$ denotes the fraction of the grain surface area that is substantially exposed to the stream of particles.  

From Equations (\ref{eq:GamMAT}), we see that an arbitrary grain shape of $N_{\rm fc}$ facets  has mechanical torques reduced by a factor $\sqrt{N_{\rm fc}}$. Such a suppression arises from averaging individual torques of random facets, which we term {\it cancellation effect}. For instance, a spheroidal or spherical shape of $N_{\rm fc}\gg 1$ experience negligible mechanical torques, as expected. 

The rotation frequency achieved after a time $t$ is given by
\bea
\Omega &=& \frac{\Gamma_{\rm MAT}t}{I_{1}}\nonumber\\
&\simeq& 10^{-5}\left(\frac{s_{d}}{10}\right)^{2}\left(\frac{n_{\rm H}}{30\cm^{-3}}\right)\left(\frac{1\rm km}{R}\right)^{2}\frac{t_{\rm Gyr}}{\alpha_{1}N_{\rm fc}^{1/2}}~{\rm rad \s^{-1}},~~~~\label{eq:omega}
\ena
where $s_{d}=v_{d}/v_{\rm th}$ with $v_{\rm th}=(2k_{B}T_{\gas}/m_{\rm H})^{1/2}$, and the travel time $t_{\rm Gyr}$ is expressed in units of $10^{9}$ yr. Here $\gamma_{e}=1/6$ is assumed. For perfect reflection, $\gamma_{r}=1$, and sticking collisions have the torque reduced by a factor of 2.

The rotation period induced by mechanical torques is
\bea
P_{\rm MAT}\simeq 175s_{d,1}^{-2}R_{\km}^{2}\left(\frac{30\cm^{-3}}{n_{\rm H}}\right)\frac{\alpha_{1}N_{\rm fc}^{1/2}}{t_{\rm Gyr}}~{\rm hr},~\label{eq:PMAT}
\ena
where $s_{d,1}=(s_{d}/10)$, and $R_{\km}$ is the asteroid radius in km. 

For an irregular shape with high helicity, e.g., $N_{\rm fc}\sim 10$, a small asteroid of $R=0.1$ km moving at supersonic speed ($s_{d}=10$) can be spun-up to $P_{\rm MAT}\approx 0.55$ hr after $t=10$ Gyr. Note that the minimum rotation period of large SASs is $P_{\min}\sim 2.2$ hr \citep{2000Icar..148...12P}.

Above, the gas-asteroid collisions are treated using the kinetic approach, which is valid for the ISM. When the IAS encounters a very dense region such as a protoplanetary disk with gas density $n_{\rm H}\sim 10^{11}\cm^{-3}$, the fluid approach should be used because the mean free path of gas atoms is $\lambda_{\rm mfp}\sim 1/n_{\H}\sigma\sim 1/(10^{11}\cm^{-3}10^{-15}\cm^{2})\sim 0.1$ km, comparable to the ISA's radius under our interest in this paper.

\subsection{Spinup by impulsive torques from Dust collisions}
Collisions with interstellar dust grains can provide an ISA some impulsive torques. For a typical interstellar medium with the dust-to-gas mass ratio of $0.01$, the mean distance between two dust grains is roughly $0.1$ km, assuming for simplicity that all dust mass is contained in $0.1\mu$m grains. Therefore, we can assume that the collisions with dust particles can be considered discretely, resulting in impulsive torques, as shown for the case of interstellar spacecraft \citep{Hoang:2017hg}. 

A single collision with a dust grain of mass $m_{\rm gr}$ can deposit an angular momentum on the small asteroid over a very short interaction time:
\bea
\delta J \sim \frac{m_{\rm gr}v_{d}R}{2},\label{eq:dL}
\ena
which corresponds to the increment of angular velocity
\bea
\delta \omega =\frac{\delta J}{I_{1}} \approx \frac{15m_{\rm gr}v_{d}}{4\pi \rho \alpha_{1}R^{4}},~~~~\label{eq:domega}
\ena
where $\rho$ is the dust mass density.

The total increase of the asteroid angular momentum after experiencing $N_{\rm coll}$ collisions can be estimated using the random walk formula:
\bea
\Delta J = N_{\rm coll}^{1/2}\delta J.\label{eq:deltaJ}
\ena
Assuming that all interstellar dust mass is in $0.1\mum$ particles and a dust-to-gas mass ratio of 0.01, the total number of collisions with $0.1\mum$ grains is
\bea
N_{\rm coll} &\sim & \frac{0.01v_{d}t m_{\H}\pi R^{2}}{(4\pi/3)\rho \times 10^{-15}\cm^{3}}\nonumber\\
&\sim& 10^{21}s_{d,1}R_{\km}^{2} t_{\rm Gyr}.~~~\label{eq:Ncoll}
\ena

The rotation frequency after $N_{\rm coll}$ (Eq. \ref{eq:Ncoll}) collisions can now be calculated using the total angular momentum $\Delta J$ from Equation (\ref{eq:deltaJ}):
\bea
\Omega &\sim& \frac{\Delta J}{I_{1}}\sim \sqrt{N_{\rm coll}}\delta \omega\nonumber\\
&\simeq& 10^{-17}\left(\frac{s_{d,1}^{3/2}}{\alpha_{1}R_{\km}^{3}}\right)t_{\rm Gyr}^{1/2}a_{-5}^{3}{\rm rad}\s^{-1},
\ena
which is negligible compared to the spinup by mechanical torques. This is due to the random walk which suppresses the dust effect by a factor of $N_{\rm coll}^{1/2}$.

Note that dust collisions can erode the surface of the asteroid. Yet we expect this to be a minor effect due to the large volume of ISAs.

\subsection{Rotational Damping and ISA Slowing Down}
The spinup is accompanied by rotational damping and slowing down by interstellar gas. 

The rotation of an ISA undergoes damping due to the sticking collisions of gas atoms to the asteroid surface and by evaporation of thermalized molecules. The damping time is equal to
\bea
\tau_{\gas}&=&\frac{3I_{1}}{4\sqrt{\pi}n_{\H}m_{\H}v_{\rm th}R^{4}}\nonumber\\
&\simeq & 1.5\times 10^{6}\alpha_{1}R_{\rm km}\left(\frac{300\cm^{-3}\K^{1/2}}{n_{\H}T_{\gas}^{1/2}}\right) ~{\rm Gyr}.\label{eq:tgas}
\ena 
Thus, the rotational damping of ISA rotation by gas collisions is much lower than the spinup time.

The slowing down time of ISAs due to gas collisions is also long
\bea
\tau_{\rm slow}&=&\frac{Mv}{F_{\rm drag}}=\frac{Mv}{n_{\H}m_{\H}v^{2}R^{2}}\sim \frac{R}{n_{\H}m_{\H}v}\\
&=& 10^{5}s_{d,1}^{-1}R_{\rm km}~ {\rm Gyr},\label{eq:tslow}
\ena
where we have assumed that ambient gas particles transfer their entire momentum upon collisions. If they are simply reflected, this increases the drag force by a factor of 2. 

Thus, ISAs travel a long distance in the ISM until it is rotationally disrupted or destroyed by collision with interstellar objects.

\subsection{Inelastic Relaxation and Wobbling of ISAs}
An isolated ISA spinning along a non-principal axis undergoes internal dissipation due to imperfect elasticity. The material within a spinning body is stressed by centrifugal forces. When the rotation is not along a principal axis, the angular velocity precesses around the axis of maximum moment of inertia. Thus, the stress at a point in the asteroid has a rotational component with respect to the asteroid, which results in the dissipation of the rotation energy into heat. This internal relaxation eventually leads to the perfect alignment of the axis of major inertia with the angular momentum with the minimum rotational energy (\citealt{1973MNRAS.165..403B}; \citealt{1979ApJ...231..404P}). 

\cite{1973MNRAS.165..403B} derived the characteristic timescale of internal relaxation due imperfect elasticity: 
\bea
\tau_{\rm rel} &=&D\frac{\mu Q}{\rho R^{2}\Omega^{3}}\nonumber\\
&\simeq & 0.02 \mu_{12}\left(\frac{Q}{10^{3}}\right)\left(\frac{D}{100}\right)\left(\frac{P_{\rm hr}^{3}}{R_{\km}^{2}}\right)~~\rm Gyr, 
\ena
where $\mu$ is the asteroid rigidity with $\mu_{12}=(\mu/10^{12} {~\rm dyne}\cm^{-2})$, $Q$ is the anelasticity of the material, $D$ is the scaling constant, and $P_{\rm hr}$ is the rotation period in hours.

The $D$ value for asteroids is uncertain. \cite{1973MNRAS.165..403B} estimated $D\sim 100$ for nearly spherical objects and $D\sim 20$ for elongated objects. \cite{Efroimsky:2000p5384} revisited the inelastic relaxation and found that the relaxation rate can be increased by an order of magnitudes, which show $D\sim 5$ for elongated objects. A recent study by \cite{2005MNRAS.359...79S} obtained even a larger value of $D\sim 100-200$. More recent revisits are presented in \cite{2012MNRAS.427..755B} for irregular asteroids and \cite{2018MNRAS.473..728F} for oblate asteroids. In general, the internal relaxation is long compared to the breakup timescale by mechanical torques for highly irregular grains with $R< 1$ km and period of several hours.

\section{Rotational Disruption}\label{sec:breakup}
\subsection{Disruption by Mechanical Torques}
A rigid body of radius $R$ rotating at velocity $\Omega$ develops a stress which scales as $S=\rho \Omega^{2}R^{2}/4$. Let $S_{\max}$ be the maximum stress of the material. Then, the critical limit for the rotational breakup is given by (\citealt{1979ApJ...231..438D}):
\bea
\Omega_{\rm cr} &\sim & \frac{2}{R}\left(\frac{S_{\max}}{\rho} \right)^{1/2}\nonumber\\
&\sim& 3.6\times 10^{-4}R_{\km}^{-1}\left(\frac{S_{\max}}{10^{3} \erg\cm^{-3}}\right)^{1/2} {\rm rad \s^{-1}}.~~~\label{eq:omega_cr}
\ena

The maximum stress for ISAs is uncertain. Numerous studies \citep{2000Icar..148...12P} find that large SSAs are usually slow rotators, while {asteroids smaller than 0.1 km of radius} are fast rotators.\footnote{Note that several large SSAs as super-fast rotators are recently discovered (see \citealt{2017GSL.....4...17C}).} In addition, the rotation rate of small SSAs tends to increase with radius, while large SSAs have a upper limit of rotation rate at a period of $P\sim 2.2$ hr. Thus, small asteroids have constant large stress $S_{\max}$, whereas large asteroids are expected to have low $S_{\max}$ (e.g., rubble-pile asteroids). For large SSAs of rubble pile, $S_{\max}\sim 10^{2}-10^{3} \erg\cm^{-3}$. For {small asteroids} ($R<1$ km), based on results from \cite{2000Icar..148...12P}, we estimate the strength $S_{\max}\sim 10^{3}\erg\cm^{-3}$ for $R=0.5$ km. Thus, we take this as a typical value {in the following calculations} for small asteroids (\citealt{2002aste.book..463A}).

From Equations (\ref{eq:omega}) and (\ref{eq:omega_cr}), we obtain the characteristic timescale to reach the first breakup is 
\bea
\tau_{\rm br}\sim 10R_{\km}\left(\frac{30\cm^{-3}}{n_{\rm H}}\right)s_{d,1}^{-2}S_{3}^{1/2}\alpha_{1}N_{\rm fc,1}^{1/2} ~{\rm Gyr},\label{eq:tcr}
\ena
where $S_{3}=(S/10^{3}\erg\cm^{-3})$, and $N_{\rm fc,1}=(N_{\rm fc}/10)$.

\begin{figure*}
\centering
\includegraphics[width=0.45\textwidth]{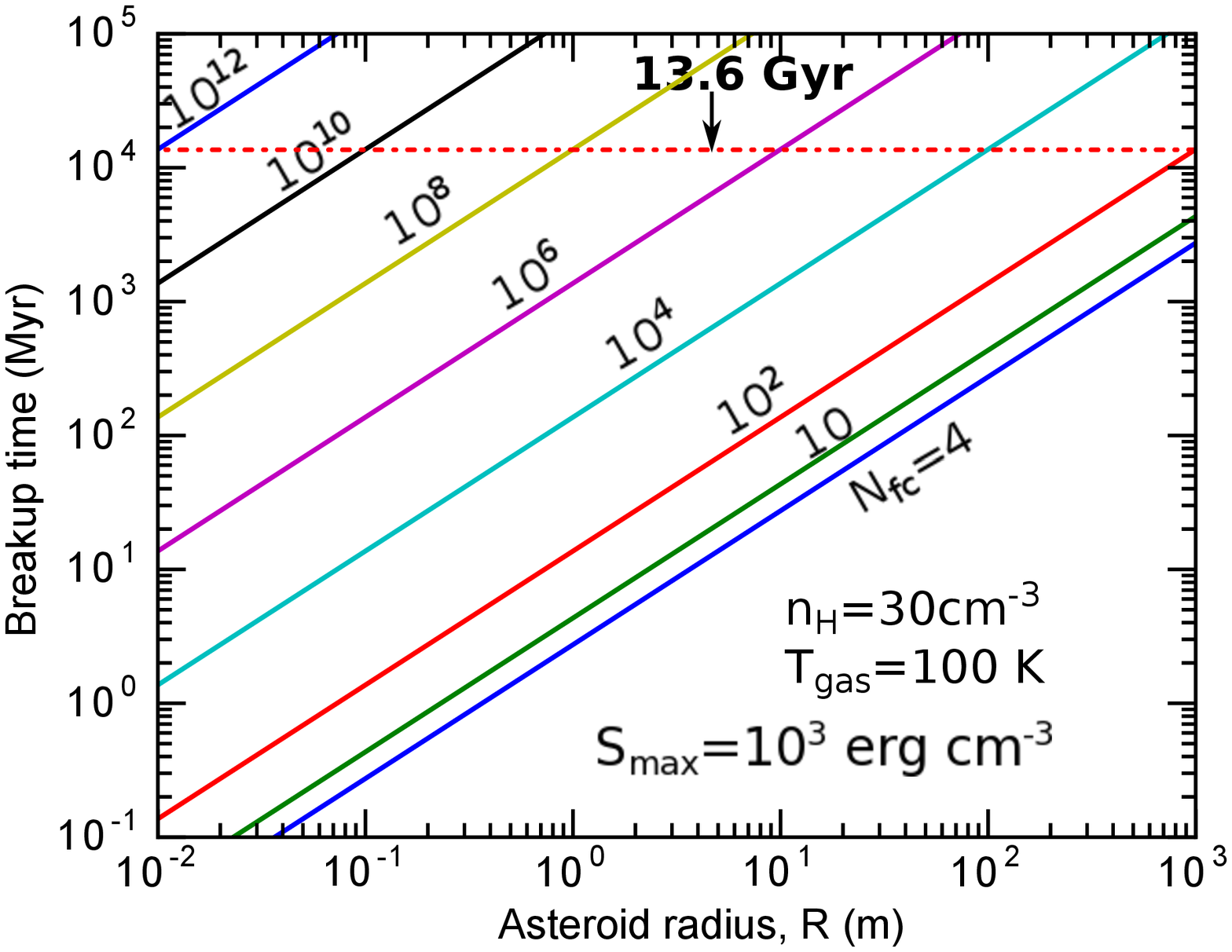}
\includegraphics[width=0.45\textwidth]{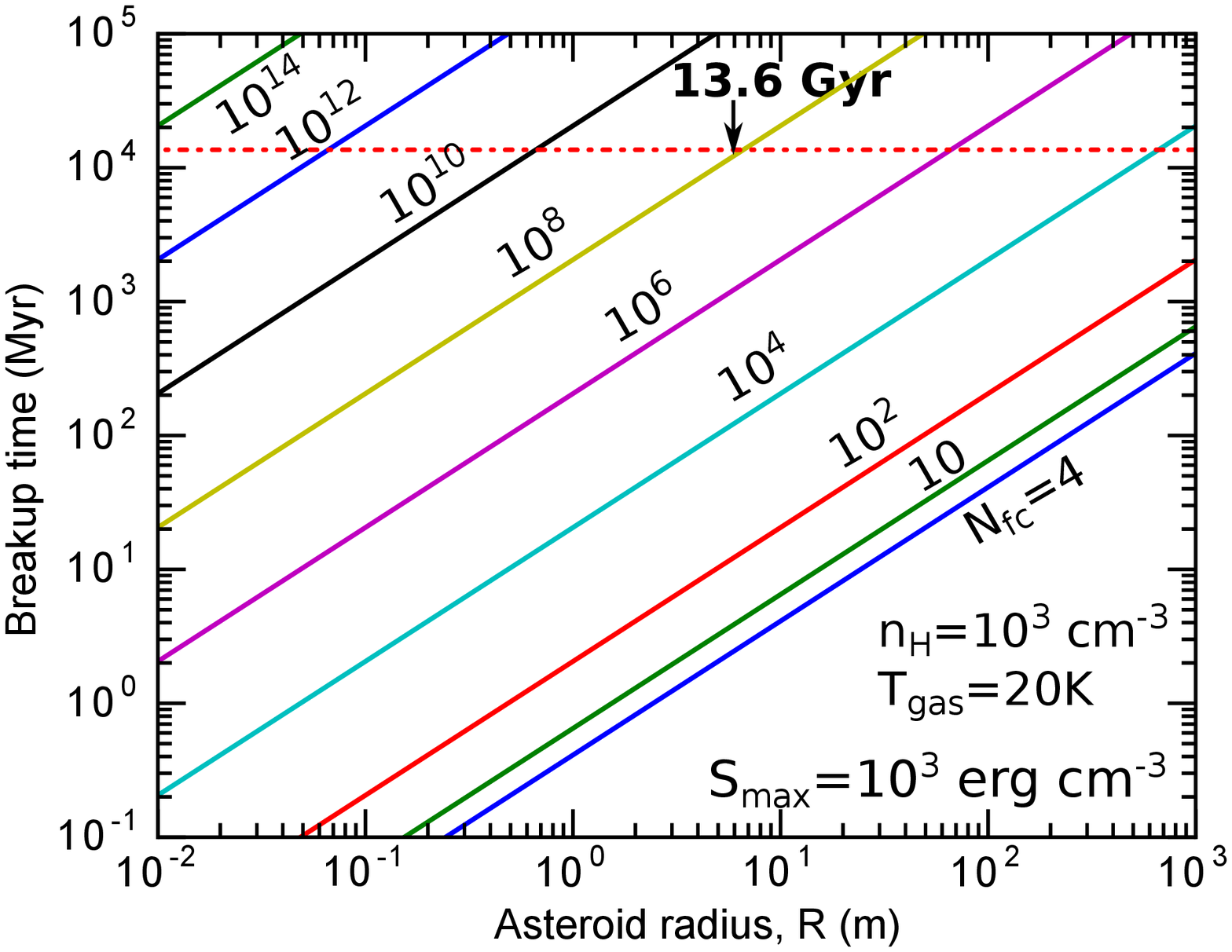}
\caption{Characteristic timescale of rotational breakup by mechanical torques vs. radius of ISA for material of maximum stress $S_{\max}=10^{3} \erg\cm^{-3}$, assuming $s_{d}=10$. Different values of the asteroid shape parameter $N_{\rm fc}$ are considered, from $N_{\rm fc}=4$ to $10^{14}$. The horizontal like shows the maximum age of stars in the Milky Way of 13.6 Gyr. Left and right panels show results for the characteristic gas density and the gas temperature of the diffuse ISM and a dense interstellar cloud, respectively.}
\label{fig:taubr}
\end{figure*}

Figure \ref{fig:taubr} (left panel) shows the lifetime of ISAs as a function of size for the diffuse ISM conditions ($n_{\H}=30\cm^{-3}, T_{\gas}=100$ K), assuming a maximum strength $S_{\max}=10^{3}\erg\cm^{-3}$ and a typical speed of $s_{d}=10$. The breakup time increases rapidly with increasing $R$, indicating that small asteroids tend to be shorter lived than large asteroids in the ISM. This is because the big asteroid has a larger moment of inertia that requires a long time to spinup to $\Omega_{\rm cr}$. For highly asymmetric shapes, the breakup time is rather short, but it is extremely long for symmetric shape with $N_{\rm fc}\sim 10^{14}$. For the chosen strength $S_{\max}$, even a large asteroid of $R\sim 1$ km can be disrupted within $10$ Gyr by mechanical torques, given a highly irregular shape. 

We note the ISA lifetime scales as $1/n_{\rm H}$. Thus, when the ISA encounters a dense molecular cloud in the ISM, its lifetime is significantly decreased. The right panel of Figure \ref{fig:taubr} shows the results for a typical giant molecular cloud of $n_{\rm H}=10^{3}\cm^{-3}, T_{\gas}=20$ K and $D=1$ pc, requiring a period of $T=(D/v_{d}) \sim 10^{18}/10^{5} \sim 1$ Myr for a ISA to cross the cloud.
 
\subsection{Upper Limits on ISA's speed}
A related effect is the variation in the velocity of fragments due to rotational breakup. Indeed, the spinup process due to mechanical torques acts to convert the kinetic energy of the original ISA into the kinetic energy of secondary ISAs. The secondary ISAs continue to move into the ISM, are spun-up by mechanical torques, and be disrupted again. Let us consider a simple model in which an original asteroid (e.g., {\it mother asteroid}) is instantaneously disrupted into two fragments of equal volume when $\omega$ reaches the critical value. The radius of the fragment after $n$ breakups is calculated as
\bea
r_{n}=\frac{R}{a^{n}},~~{\rm with~} a=2^{1/3},\label{eq:rn}
\ena
where $r_{0}=R$.

The launch speed of a fragment with respect to the mother's center of mass is approximately equal to the tangential velocity
\bea
\delta v_{n}=\Omega_{\rm cr}r_{n} \sim 0.0002 S_{3}^{1/2} ~{\rm km} \s^{-1},\label{eq:v1}
\ena
where $\Omega_{\rm cr}$ is given by Equation (\ref{eq:omega_cr}), which depends only on the maximum stress of the material. For a typical strength of $S_{\max}=10^{3} \erg\cm^{-3}$, $\delta v_{n}$ is rather small, but it can reach $\delta v_{n}\sim 2 {\rm km}\s^{-1}$ for a strong material with $S_{\max}\sim 10^{11} \erg\cm^{-3}$ such as single-crystalline. 

Due to the increase of mechanical torques with the speed, asteroids moving extremely fast would be rapidly disrupted by rotational breakup. Thus, the rotational disruption can constrain the upper limit for ISA speeds. The critical speed for an ISA with the lifetime $t$ is given by
\bea
s_{d,1}^{2}= \left(\frac{10}{t_{\rm Gyr}}\right)\left(\frac{30\cm^{-3}}{n_{\rm H}}\right)S_{3}^{1/2}\alpha_{1}N_{\rm fc,1}^{1/2}R_{\km},~~\label{eq:sd_cr}
\ena
which depends on the size, shape, and strength of the asteroid. For an asteroid of $R \sim 1$ km and lifetime $t=1$ Gyr, the maximum speed is $s_{d}\sim 20$, or 20 km$\s^{-1}$ in the typical diffuse ISM, assuming a highly irregular shape of $N_{\rm fc}=4$.

\subsection{Maximum distance traveled of ISAs}
Let us suppose that an asteroid is produced in a stellar system within our Galaxy at distance $D$ from the Earth. Due to rotational breakup, ISAs from very distant stellar systems would be disrupted before reaching the Earth.  The maximum distance from where an asteroid can reach us is given by
\bea
D_{\max} &=& v \tau_{\rm br}\nonumber\\
&\sim& 10R_{\km}\left(\frac{30\cm^{-3}}{n_{\rm H}}\right)s_{d,1}^{-1}S_{3}^{1/2}\alpha_{1}N_{\rm fc,1}^{1/2} ~{\rm kpc}.\label{eq:Dmax}
\ena

\begin{figure*}
\centering
\includegraphics[width=0.45\textwidth]{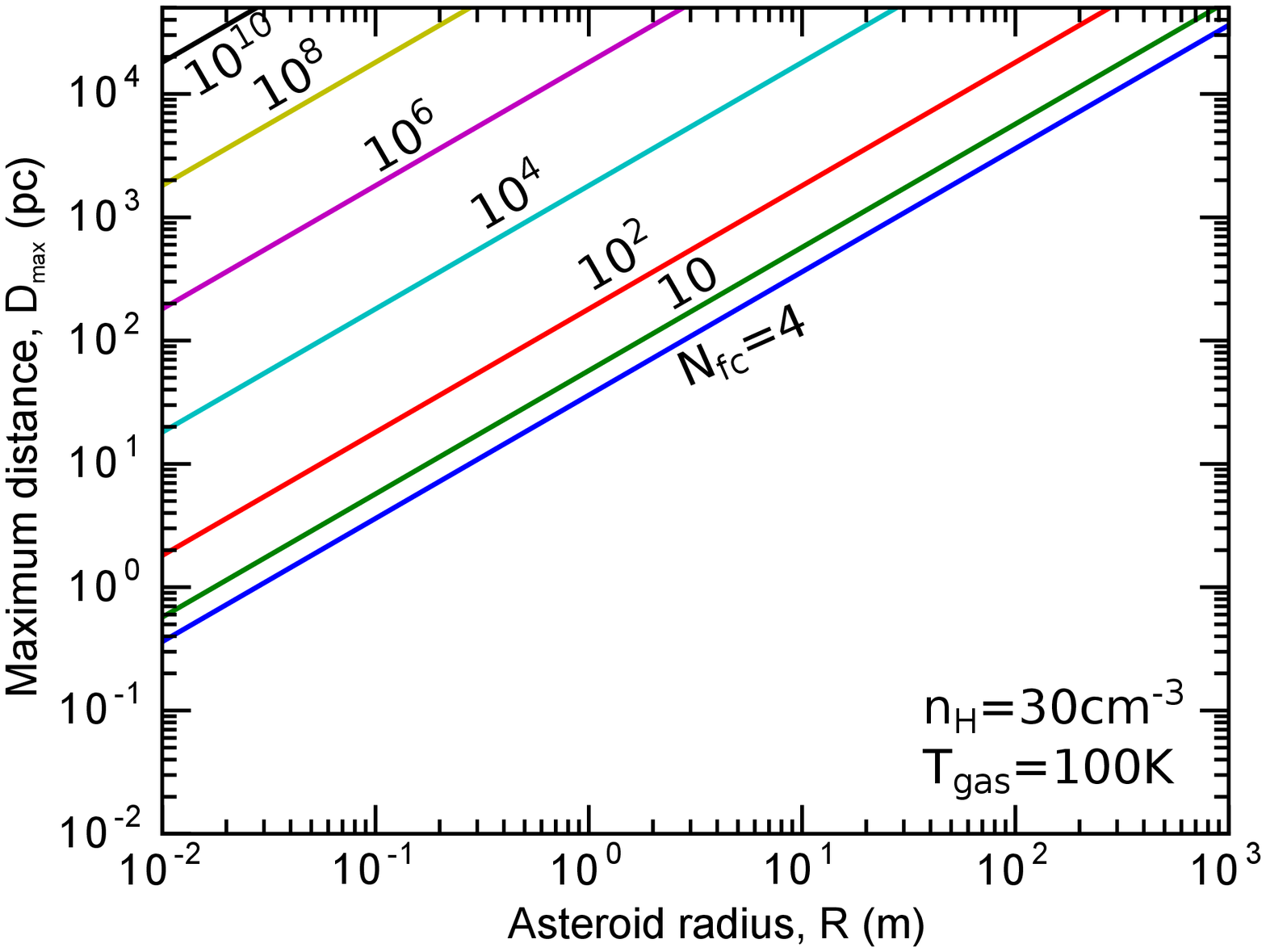}
\includegraphics[width=0.45\textwidth]{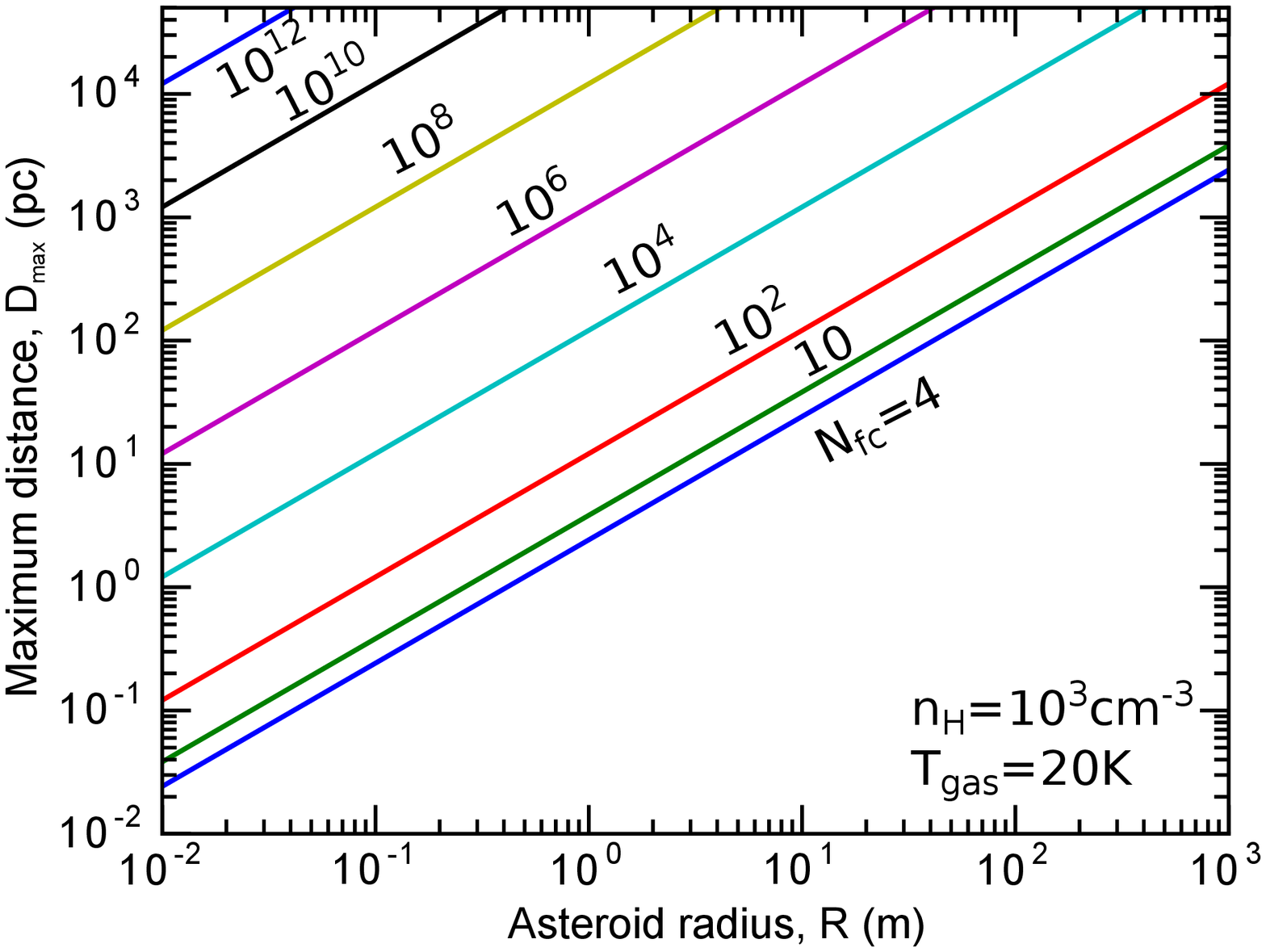}
\caption{Same as Figure \ref{fig:taubr}, but for the maximum distance from where the ISA could reach Earth without being disrupted by mechanical torques.}
\label{fig:Dmax}
\end{figure*}

Figure \ref{fig:Dmax} shows the maximum distance over which an asteroid ejected could survive its journey in the diffuse ISM (left panel) and a dense molecular cloud (right panel) to visit us, assuming a typical speed of $s_{d}\sim 10$.
Asteroids with highly irregular shapes originate from a distance $D> D_{\max}\sim 1$ kpc would be disrupted.

\section{Discussion}\label{sec:discuss}
\subsection{Dynamics of interstellar asteroids and its effects}
The dynamics of SSAs is well studied in the literature. Violent collisions with asteroids, the YORP effect, and tidal encounters are the three main mechanisms that affects the dynamics of SSAs. In particular, the rotational disruption by the YORP effect is a leading mechanism for the formation of a binary SSAs (\citealt{2007ApJ...659L..57C}; \citealt{2008Natur.454..188W}). The former process is expected to be not important for ISAs because the number density of ISAs is much lower than in the solar system. The interstellar radiation field, although having some degree of anisotropy, is several orders of magnitude lower than sunlight in the Solar system, making the interstellar YORP effect inefficient (\citealt{Gaidos:2017wj}). 

In this paper, we extend previous works on mechanical torques formulated for dust grains (\citealt{2007ApJ...669L..77L}; \citealt{2018ApJ...852..129H}) to ISAs, and study the dynamics of ISAs subject to this torque. By modeling the irregular shape by a number of flat facets $N_{\rm fc}$ of random orientation, we calculate the mechanical torques for ISAs and find that the torques can spinup small ISAs to high rotation speeds, resulting in the disruption of IASs into small fragments. The breakup by mechanical torques depends on the size, shape, as well as the strength of the asteroid material. 

The rotational disruption by mechanical torques can be a dominant mechanism for the formation of binary asteroids. The resulting fragments would wobble around the original spin axis. The binary asteroids produced by rotational disruption should have similar spin orientations. 

\subsection{Expected Shape of Interstellar Asteroids}
ISAs are expected to have a variety of initial irregular shapes, similar to SSAs. While it is widely believed that collisions can smooth the sharp edges of asteroids, several SSAs exhibit simple shapes with large planar facets and sharp edges (\citealt{Torppa:2003ii}), such as 2100 Ra-Shalom, 43 Ariadne, 694 Ekard (see \citealt{2009ApJ...699L..13D}).

The breakup time (Eq. \ref{eq:tcr}) and maximum travel distance (Eq. \ref{eq:Dmax}) are derived assuming a fixed ISA geometry. Under realistic conditions, the shape of ISAs changes gradually over time due to micro-collisions with dust grains (\citealt{2009ApJ...699L..13D}). The process is suggested to form asteroids with rounder surfaces, or more simple shapes with planar facets and sharp edges \citep{2009ApJ...699L..13D}. {The asteroid elongation may be increased by collisions with a large number of small projectiles (\citealt{2015MNRAS.454.1704H})}. However, without collisions with other asteroids, the timescale for significant change in the shape may be longer than the spinup time by mechanical torques.

Our results were derived for the case of perfect reflection and irregular shape of the ISA. Yet the surface of ISAs is expected to be non-uniform due bombardment of interstellar dust grains and cosmic rays. As a result, an asteroid of highly symmetric shape may still acquire considerable mechanical torques due to the variation in reflectivity over its surface.

\subsection{Implications for `Oumuamua}
`Oumuamua is the first interstellar asteroid detected to date, {which is suggested to be a rocky or devolatilised object \citep{Jackson:2017tq}, or a cometary nucleus covered by an insulating mantle \citep{Fitzsimmons:2017io}}. The discovery of `Oumuamua is not expected as the fraction of asteroids was estimated to be between $10^{-4}-10^{-2}$, i.e., much lower than cometary objects (\citealt{2009ApJ...704..733M}). This may be due to the fact that cometary objects are more efficiently disrupted by mechanical torques. 

We can estimate the lifetime of `Oumuamua to be $t_{\rm br}\sim 0.5$ Gyr using Figure \ref{fig:taubr} for $R\sim 200$ m, and the highest irregularity $N_{\rm fc}\sim 4$. Thus, `Oumuamua was possibly a product of breakup rather than an original asteroid formed several billions year ago  (see also \citealt{Jackson:2017tq}). 

The elongated shape of `Oumuamua is a puzzle because no such elongated shape is found for SSAs. For SSAs, the YORP effect is expected to produce strange shapes. \cite{Domokos:2017tn} suggested the abrasion of a large number of dust grains can explain that shape. Here, we suggest that `Oumuamua may be formed via a process, involving spinup, disruption, and recombination. First, the original asteroid is ejected from a stellar system and is rapidly moving in the ISM. During its motion, it is spun-up to high speeds by mechanical torques, resulting in an increase in the elongation. When the rotation rate reaches the critical limit, the original asteroid breaks to binary fragments. The binary fragments continue to be spun-up by mechanical torques and become more elongated. {Subsequent evolution may induce the reassembly of the fragments to form a contact binary asteroids which might have extreme elongation.}\footnote{The physics underlying the formation of contact binary asteroids is not clear, including binary YORP effect, gravity combined with spin-down by YORP, and low-velocity collisions of fragments produced by secondary disruption (see \citealt{2007Icar..189..370S}; \citealt{2014Icar..229..418T}; \citealt{2011Icar..214..161J}). The YORP and BYORP effects would not work in the ISM due to the diffuse radiation field, whereas the secondary disruption cannot produce a contact binary with similarly sized components.} Interestingly, mechanical torques can cause the spin-down of binary asteroids when the axis of maximum moment of inertia is at an oblique angle from the direction of the motion (\citealt{2018ApJ...852..129H}). Thus, the spin-down by mechanical torques appears to be a plausible mechanism to form interstellar contact binary asteroids.

Let us estimate if the mutual gravity is strong enough to bind the fragments against the centrifugal force. The gravity of two equal mass fragments is $F_{\rm gra}= GM^{2}/R^{2} \sim 6\times 10^{12}$ dyn for $R=0.1$ km and $M\sim 10^{14}(R/0.1 \rm km)^{3}$ g. The centrifugal force at the current rotation period is $F_{\rm cf}= M(2\pi/P)^{2} R \sim 10^{11} (8 \rm hr/P)^{2}$ dyn. Therefore, gravity is sufficiently strong to keep these fragments rotating at $P\sim 8$ hr.

A more exotic scenario is that `Oumuamua is composed of a central core object of radius $R$ that carries most of the mass at rock density $\rho$, surrounded by a disk of small fragments. Thus, the Newtonian rotation period of low-mass fragments around the core would be, $P_{\rm obs}=\left(3\pi/G\rho\right)^{1/2}\left(R_{\rm obs}/R \right)^{3/2} \sim 8 (R_{\rm obs}/3R)^{3/2}$ hr, for $\rho=4.5\g\cm^{-3}$. This matches the spin period of `Oumuamua for $R_{\rm obs}\sim 3R$. The variation in projected area would be $\sim (R_{\rm obs}/R)^{2}\sim 10$, also similar to the observed flux variation amplitude for `Oumuamua.
  
Finally, `Oumuamua is found to be spinning along the non-principal axis \citep{Fraser:2018dg}. Based on the estimate of the internal relaxation time being billion years, \cite{Fraser:2018dg} suggested that the tumbling was produced in the stellar system before the object was ejected to the ISM. In light of our study, the tumbling can be induced by rotational disruption due to mechanical torques. {Indeed, mechanical torques induced by gas bombardment on an irregular asteroid have three components, one component parallel and two perpendicular to the principal axis of maximum moment of inertia (\citealt{2007ApJ...669L..77L}; \citealt{2018ApJ...852..129H}). These perpendicular components induce rotation along the two principal axes, i.e., the tumbling of the asteroid. Note that no tumblers formed by rotational disruption are observed among solar system asteroids \citep{2018Icar..304..110P}. This suggests that the disruption mechanism may be different from mechanical torques.}

\subsection{Implication for high-velocity asteroids}
Observations show the existence of a population of high-velocity stars, including runaway stars, hypervelocity stars (HVSs) with velocity $v\ge 10^{3} \km \s^{-1}$. The latter is expected to be produced by the disruption of stellar binaries by a black hole at our Galactic center (\citealt{1988Natur.331..687H}; \citealt{2015ApJ...806..124G}).  High-velocity stars are expected to have their own planetary system with asteroids and comets (\citealt{2009A&A...496..307K}; \citealt{2012MNRAS.421.1315Z}; \citealt{2012MNRAS.423..948G}). Therefore, the formation of HVSs would also produce {\it high-velocity asteroids} (thereafter HVAs) moving out of the Galactic center. The detection of a HVA is important for testing the idea of planet formation of stars orbiting a black hole \citep{2012MNRAS.419.1238N}.
 
Let us estimate the maximum speed of HVAs that can travel to us. By setting $D_{\max}$ to the distance from Galactic center, $D_{\rm GC}\sim 8$ kpc, we can derive the maximum speed that can survive as
\bea
s_{d,\max}\sim  1.3\left(\frac{8\rm kpc}{D_{\rm GC}}\right)\left(\frac{30\cm^{-3}}{n_{\rm H}}\right)R_{\km}S_{3}^{1/2}\alpha_{1}N_{\rm fc,1}^{1/2}.~~~\label{eq:sd_max}
\ena

Thus, HVAs of highly irregular shape would be disrupted before reaching the Earth, but those of highly symmetric shape can survive. Note that HVAs would be gradually evaporated due to heating by interstellar gas, which can lower their detectability
 
\section{Summary}\label{sec:sum}
In this paper, we have studied the effect of mechanical torques due to drifting of interstellar asteroids in the ISM. We found that ISAs of irregular shapes can be spun-up by mechanical torques when moving in the ISM. The spinup efficiency is strongest for asteroids with several flat surfaces and clear-cuts, which are not unusual for SSAs.

We showed that rapidly spinning ISAs can be disrupted into small fragments when the rotation rate exceeds the maximum stress of the material. The breakup time is short, of $1$ Gyr for highly irregular shapes, and increases rapidly with the asteroid size. This might be the main mechanism resulting in the fragmentation of large asteroids and the formation of small binary asteroids in the ISM. This mechanism is also important for the destruction of interstellar contact binary asteroids. As a result, the extreme elongation of `Oumuamua may originate from the reassembly of the binary fragments previously formed by rotational disruption due to mechanical torques.

We estimated the internal relaxation of ISAs due to inelastic viscosity effect. The relaxation time may be longer than the breakup time. The tumbling of `Oumuamua can result from rotational disruption due to mechanical torques when the asteroid was moving through the ISM.

We discussed the chance of detecting high-velocity asteroids presumably formed in the high-velocity stellar system near the Galactic center. We find that the highly irregular asteroids are likely disrupted before reaching the Earth, but highly symmetric HVAs would reach us.

Our results should also apply to Oort cloud asteroids which are outside the heliosphere and have exposed to the interstellar conditions.

\acknowledgements
We thank an anonymous referee for insightful comments. T.H. acknowledges the support by the Basic Science Research Program through the National Research Foundation of Korea (NRF), funded by the Ministry of Education (2017R1D1A1B03035359). A.Loeb was supported in part by the Breakthrough Prize Foundation. We thank E.Turner for elucidating discussions.

\bibliography{ms.bbl}

\end{document}